\documentstyle{article}
\def\tilde{\widetilde}
\def\bar{\overline}

\def\*{\star}
\def\({\left(}          
\def\){\right)}         
\def\[{\left[}          
\def\]{\right]}

\def\frac#1#2{{#1 \over #2}}

\def\d{\partial}

\def\2pi{\hbox{$2\pi i$}}

\def\dsl{\raise.15ex\hbox{/}\kern-.57em\partial}
\def\Dsl{\,\raise.15ex\hbox{/}\mkern-.13.5mu D}
\def\ga{\gamma}         

\def\al{\alpha}

\def\la{\lambda}        \def\La{\Lambda}

\def\CA{{\cal A}}              \def\CC{{\cal C}}

              \def\CO{{\cal O}}
\def\CP{{\cal P}}       \def\CQ{{\cal Q}}

\def\debut{ \begin{eqnarray} }
\def\fin{ \end{eqnarray} }
\def\non{ \nonumber }

\def\presentation{
\voffset -.50in
\hoffset -.19in
\oddsidemargin 0in \evensidemargin 0in
\marginparwidth .75in \marginparsep 7pt \topmargin 0in
\headheight 12pt \headsep .25in
\footheight 18pt \footskip .35in
\textheight 9.5in \textwidth 6.5in
\columnsep 10pt \columnseprule 0pt }

\presentation
\begin{document}

\rightline{LPTHE-98-24 }
\vskip 1cm
\centerline{\LARGE Structure of Matrix Elements in Quantum Toda Chain. }
%\bigskip
%\centerline{\LARGE the Restricted Sine-Gordon Model.}
\vskip 2cm
\centerline{\large F.A. Smirnov
\footnote[0]{Membre du CNRS}
\footnote[1]{On leave from Steklov Mathematical Institute,
Fontanka 27, St. Petersburg, 191011, Russia} }
\vskip1cm
\centerline{ Laboratoire de Physique Th\'eorique et Hautes
Energies \footnote[2]{\it Laboratoire associ\'e au CNRS.}}
\centerline{ Universit\'e Pierre et Marie Curie, Tour 16 1$^{er}$
		\'etage, 4 place Jussieu}
\centerline{75252 Paris Cedex 05 France}
\vskip2cm
{\bf Abstract.} We consider the quantum Toda chain using the
method of separation of variables. We show that the matrix elements of
operators in the model are written in terms of finite number
of ``deformed Abelian integrals''. The properties of these integrals
are discussed. We explain that these properties 
are necessary in order to
provide the correct
number of independent operators. The comparison with the
classical theory is done.
\newpage

\section{Introduction}

As it became clear recently \cite{bbs,bbs1,cft} there is
a close connection between the formulae for the
matrix elements in integrable field theory (form factors)
\cite{book} and
the method of separation of variables developed by Sklyanin \cite{skl}.

The form factors are typically given by certain integrals. These
kind of
formulae can be interpreted as follows. Consider an integrable model
which allows the separation of variables. The separated variables
naturally split into two equal parts: one of them can be
considered as ``coordinates'' and another as ``momenta''
(of course they have nothing to do with original canonical
variables in which the model is formulated).
The formulae for the form factors are understood as
matrix elements written in ``coordinate'' representation, i.e. in
terms of integrals with respect to the ``coordinates''.

Another observation made in \cite{dai}, and used intensively in 
\cite{bbs,bbs1} is that the integrals in the formulae for
the form factors 
in models with $\widehat{sl}(2)$ Lie-Poisson symmetry
(Sine-Gordon, for example)
can be considered as deformations of hyper-elliptic
integrals. This fact must be also related to the method
of separation of variables because the ``coordinates'' describe
classically a divisor on the spectral hyper-elliptic curve.
The important conclusion made in \cite{bbs1}
is that these deformed hyper-elliptic integrals
must have similar properties to the usual hyper-elliptic
integral in order that the correct number of equations of
motion exists in the quantum case.

In paper \cite{cft} we performed the quasi-classical analysis
of the matrix elements in CFT in finite volume. This is a much more
complicated case than the case of infinite volume.
The method of separation of variables seems to give the only possible approach
to the calculation of form factors. The main difficulty of the
problem in finite volume is due to the fact that the separation
of variables leads to the Baxter equations whose solutions
describe the wave-functions in ``coordinate'' representation.
So, one must consider integrals of solutions over Baxter equations.

In the present paper we consider a much simpler model
which nevertheless exhibits difficulties similar
to those of integrable field theory in finite volume.
This is the periodical Toda chain. Historically this is the
first model to which the method of separation
of variables was applied \cite{skl}. In this case the problem of
describing the spectrum leads to Baxter equations with
non-trivial solutions in entire functions. The matrix elements are
given by integrals over these solutions. We show that
these integrals can be considered as
deformed hyper-elliptic integrals allowing deformation
of all the important properties of hyper-elliptic 
integrals. Similarly to \cite{bbs1} these properties are needed for
the correct counting of operators, they are actually equivalent to the 
equations of motion.

\section{Classical Toda chain}

The periodical Toda chain is described by the Hamiltonian:
\debut
H=\sum\limits _{j=1}^n {p_j^2\over 2}+e^{q_{j+1}-q_j} 
\label{ham}
\fin
where $p_j,q_j$ are canonical variables, $q_{n+1}\equiv q_1$.

The exact solution is due to existence of Lax representation.
Consider the L-operator
\debut
L_j(\la)=\pmatrix{ \la -p_j, &e^{q_j}\cr -e^{-q_j}, &0}
\non
\fin
and the monodromy matrix
\debut
M(\la)=L_n(\la)\cdots L_1(\la)=\pmatrix{A(\la)&B(\la)\cr C(\la)&D(\la)}
\non
\fin
Obviously, $\det M(\la)=1$.
The monodromy matrix satisfies Sklyanin's Poisson brackets:
\debut
\{M(\la)\ ,\hskip -0.2 cm {}^{\otimes}\  M(\mu)\}
=[r(\la -\mu), M(\la)\otimes M(\mu)]
\label{poisson}
\fin
where
$$r(\la)=-{P\over \la}$$
P is the permutation. The coefficients of $T(\la)\equiv {\rm tr} M(\la)$
are in involution:
$$\{T(\la),T(\mu)\}=0$$
Moreover,
$$T(\la)=\la ^n-P\la ^{n-1}+\({1\over 2}P^2-H\)\la ^{n-2}+\cdots$$
where $P=\sum p_j$ is the total momentum and $H$ is the Hamiltonian 
(\ref{ham}).
Thus $T(\la )$ generates $n$ integrals of motion in involution providing
complete integrability of the system.

From here on we can forget about the Toda chain saying that we consider
an orbit of Lie-Poisson group 
\cite{stsh} i.e. the polynomial matrix $M(\la)$ with 
$\det M(\la)=1$
satisfying the
Poisson brackets (\ref{poisson}) (determinant is in the center of these
Poisson brackets) and characterized by certain reality conditions
which we shall discuss later.

Let us consider the elements of $M(\la)$ in some more details.
We introduce the notations
\debut
&&A(\la)=\la ^n +\la ^{n-1}a_1+\cdots +a_n \non\\
&&B(\la)=b\(\la ^{n-1} +\la ^{n-2}b_1+\cdots +b_{n-1}\) \non\\
&&C(\la)=\la ^{n-1}c_2+\cdots +c_{n+1} \non\\
&&D(\la)=\la ^{n-2}d_2+\cdots +d_n \non
\fin
The variables $b$ and $a_1$ have the Poisson brackets
$$\{a_1,b\}=b$$
and Poisson commute with the rest of variables. In terms of Toda chain
$a_1=P$ and $b=e^{q_n}$ describe the motion of the center of mass.
Our nearest concern is the algebra of observables $\CA$.
We define this algebra as the one generated by all the monomials of 
finite degree of the variables $a_j,b_j,c_j,d_j$ and $b$. 
It is important that the polynomial structure of $M(\la)$
introduces grading of $\CA $. Namely, we can prescribe the degree
$i$ to every of elements $a_i$, $b_i$, $c_i$, $d_i$
and degree $0$ to $b$. The degrees of the leading
coefficients of $C(\la)$ and $D(\la)$ are chosen in order
that the coefficients of the determinant
$$\det M(\la)=\la ^{2n-2}f_2 +\cdots +f_{2n}$$
are homogeneous.
The variable $b$
is a kind of zero-mode, it is of minor dynamical value.
The algebra $\CA$ contains a subalgebra
$\CA _0$ of polynomial functions of
$a_i,b _i, d_i$ and $\tilde{c}_i=bc_i$.
So, this subalgebra does not have $b$ as a separate generator,
the change in definition of $c_i$ is needed in order that
the Poisson brackets are closed for
$A(\la)$, $  D(\la)$, $\tilde{B}(\la)=\prod(\la-\ga _j)$
and $\tilde{C}(\la)=bC(\la )$. We shall deal only with this
subalgebra.

The algebra $\CA _0$ as a vector space splits into direct sum of
subspaces of different degrees. Let us denote by $\delta (n)$
the dimension of 
the subspace of the degree $n$. The generating function of $\delta (n)$
(character) is given by
\debut
\chi (q)\equiv\sum\limits _{n=0}^{\infty}\delta (n)q^n=
{1\over [n]!}
\ {1\over [n-1]!}
\ {[1]\over [n+1]!}
\ {[1]\over [n]!}\ {[2n]!\over [1]}
\non%\label{char}
\fin
where $[n]=1-q^n$, 
$[n]!=[1][2]\cdots [n]$.
The first four multipliers come from monomials of
of $a_j,b_j,\tilde{c}_j,d_j$ respectively, the last
multiplier comes from factorization by the condition $\det M(\la )=1$.
%(more precisely one has to take some constant of degree $2n$ instead of 1).

Notice that
\debut
\chi (q)=
{1\over [n]!\ [n-1]!}
\(\[ 2n-1\atop n-1\]-q\[ 2n-1\atop n-2\]\)
\label{char}
\fin
where we introduced the q-binomial coefficients
$$\[ n\atop m\]=
{[n]!\over [m]!\ [n-m]!}
$$
Later we shall provide an interesting interpretation of this
formula.

Let us return to more traditional consideration of classical
Toda chain. 
We do not give a complete list of references which can be found
in \cite{skl}, the important for us fact concerning the classical system is
that it allows the separation of variables \cite{toda1,toda2}.
Consider zeros of the polynomial $B(\la)$:
$$B(\la )=b\prod\limits _{j=1}^{n-1}(\la -\ga _j)$$
and the variables $\La _j\equiv D(\ga _j)$. Notice that $\La _j =\La (\ga _j)$
where $\La (\la)$ is the eigenvalue of $M(\la )$. The variables $\ga _j$,
$\log\La _j$ are canonocally conjugated which can be shown following
\cite{skl} using (\ref{poisson}):
$$\{\ga _i,\log\La _j\}=\delta _{i,j}$$
From $\det M(\la )=1$ it follows that $A(\ga _j)=\La _j ^{-1}$.
One can reconstruct the matrix $M( \la)$ from $\ga _1,\cdots ,\ga _{n-1}$,
$\La _1,\cdots ,\La _{n-1}$, $a_1$ and $b$.
The simplectic form is written as
$$\omega =\sum\limits _{j=1}^{n-1}d\log\La _j\wedge d\ga _j
+d\log b\wedge da _1 $$
The 1-form $\al $ ($\omega =d\al $) is
\debut
\al =\sum\limits _{j=1}^{n-1}\log\La _j d\ga _j
+\log b \ da _1
\non
\fin

Let us take other coordinates on the phase space, namely, 
$\ga _1,\cdots ,\ga _{n-1}$, $t_2,\cdots ,t_n $ (defined by
$T(\la)=\la ^n+\la ^{n-1}t_1+\la ^{n-2}t_2+\cdots +t_n$), $t_1\equiv a_1$ and $b$.
From
$$\La _j=\La (\ga _j)={1\over 2}\(T(\ga _j)+\sqrt{T(\ga _j)^2-4}\)$$
one easily finds the expression for the symplectic form in
these variables
$$\omega =\sum\limits _{j=1}^{n-1}
\sum\limits_ {k=2}^n {\ga _j ^{n-k}\over \sqrt {P(\ga _j)}}\ dt _k
\wedge d\ga _j
+d\log b\wedge da _1$$
where $P(\la)=T^2(\la)-4$.
Thus the equations of motion take the form
\debut
\{T(\la),\ga _j\}&=&\sqrt{P(\ga _j)}\ \prod\limits _{k\ne j}{\la -\ga _k\over
\ga _j -\ga _k}\label{em}\cr
\{T(\la),b\}&=&\la ^{n-1}b
\fin
certainly, only the first $n-1$ equations are really interesting.
They are linearized by the Abel transformation:
$$\left\{T(\la ),\ \sum\limits _{k=1}^{n-1}
\int\limits ^{\ga _k}\sigma _j\right\}=
\la ^{j-1}$$
where $\sigma _j$ are the first kind abelian differentials
on the spectral curve $\mu ^2=P(\la)$:
$$\sigma _j={\la ^{j-1}\over \sqrt{ P(\la)}}\ d\la $$
We associate the ``times'' $\tau _1,\cdots ,\tau _{n-1}$ with
$t_2,\cdots ,t_n$:
$${\partial\over\partial\tau _j}F\equiv\partial _j F=\{t_{j+1},F\}$$
The evolution of 
$\sum\limits _{k=1}^{n-1}\int ^{\ga _k}\sigma _j $ with respect to times
is linear. 

The above considerations apply to any orbit of the Lie-poisson group.
We want now to consider specific reality conditions which
correspond to Toda chain. It can be shown \cite{toda1,toda2} that
the conditions in question are:
\newline
1. The polynomial $T(\la )$ of degree $n$ has $n$ real zeros.
Moreover its local maxima are not below $2$ and 
its local minima are 
not above $-2$.
So, all the zeros of the polynomial $P(\la)$ are also real,
they are denoted by $\la _1<\la _2<\cdots <\la _{2n}$.
\newline
2. The polynomial $B(\la )$ 
has real zeros $\ga _1, \cdots ,\ga _{n-1}$
which belong to the ``forbidden zones'': $\la _{2k}<\ga _k <\la _{2k+1}$.

The equations of motion (\ref{em}) preserve these conditions.   
The hyper-elliptic Riemann surface $\mu ^2=P(\la )$
has $2n$ branch pints ($\la _j$). Its genus equals $n-1$. We
present the surface as two complex planes with the cuts
along $(-\infty ,\la _1]$, $[\la _2,\la _3]$, ... ,$[\la _{2n}, \infty)$
identifying the banks of the cuts on two sheets in usual way.
The canonocal $a$-cycles $a_j$ are taken as ones encircling the cuts
$[\la _{2j},\la _{2j+1}]$ for $j=1, \cdots ,n-1$. Topologically the
points $\ga _j$ move along the cycles $a_j$.

Define the normalized holomorphic differentials
$$\omega _j =A_{jk}\ \sigma _k$$
such that
$${1\over 2\pi}\int _{a_j}\omega _k =\delta _{j,k}$$
Then 
$$\theta _j =\sum\limits _{k=1}^{n-1}\int ^{\ga _k}\omega _j$$
are real angles on the Jacobi variety, and the dynamics 
describes linear motion along this real torus.
One can invert the Abel transformation expressing the
symmetric functions of $\ga _1,\cdots ,\ga _{n-1}$
(recall that they coincide with $b_1,\cdots ,b_{n-1}$)
as functions on the Jacobi variety (functions of $\theta $'s)
using the
Riemann theta-function but we shall not need explicit formulae.
The angles $\theta$ and the times $\tau$ are related linearly:
$$\theta _j =\sum\limits _{l=1}^{n-1} A_{jl}\tau _{n-l}$$
so, the using the 
theta-function
formulae mentioned above one can resolve the
equations of motion 
expressing $b _ j $ as $b_j=b_j(\tau _1, \cdots ,\tau _{n-1})$.

From the point of view of algebraic geometry the monodromy
matrix $M(\la)$ gives an affine model of hyper-elliptic
Jacobian, and the functions $b _j(\tau )$ are generalized
Weierstrass functions \cite{mum}. In the case of genus one
($n=2$) the function $\ga (\tau_1)=b_1(\tau_1)$ is the usual Weierstrass
function which satisfies the second order differential equation
\debut
\partial _1^2\ga = {1\over 2}{d\over d\ga}P(\ga )
\label{weier}
\fin
One of results of our further analysis will be in finding
certain second order partial differential equations
for generalized Weierstrass functions which can be thought about
as generalizations of (\ref{weier}).

Let us consider the ring of generalized Weierstrass functions
with coefficients in $t_1,\cdots ,t_{n-1}$, i.e. the
ring of polynomials
$$F(t_1, \cdots t_n,b_1,\cdots ,b_{n-1})$$
Consider further all possible derivatives of these polynomials
with respect to $\tau _i$:
\debut
\partial _1^{k_1}\cdots \partial _{n-1}^{k_{n-1}}F(t_1, \cdots t_n,
b_1,\cdots ,b_{n-1})
\label{fun}
\fin
The equations we are looking for correspond to all possible
linear combinations of the functions (\ref{fun}) which
vanish due to equations of motion.
To understand the origin
of these equations we have to return to our mechanical considerations.

Mechanically one understands the derivatives $\partial _i$ as
action of hamiltonian vector-fields.
Using the Poisson
brackets (\ref{poisson}) one can express
(\ref{fun}) as a function of $a_1, \cdots ,a_n$, $b_1, \cdots ,b_{n-1}$,
$\tilde{c} _2,\cdots ,\tilde{c}_{n+1}$, $d_2,\cdots ,d_n$ 
i.e. as an element of the
algebra $\CA _0$. We put forward the following \newline
{\bf Conjecture 1.} {\it Every element of} $\CA _0$ 
{\it can be presented as linear combination of the expressions}
(\ref{fun}).
\newline
We were not able to find a complete proof of this statement, however,
the consideration of examples supports it. Further indirect support of
this conjecture will be provided by the calculation of characters
given below.

%One can find all this strange for the following reason.
%We are dealing with integrable system. The phase space of such system
%is foliated: the levels of integrals are tori parametrized by 
%the angle variables. The variables $t_1,\cdots ,t_n$ give complete
%set of integrals of motion. The variables $\ga _1 ,\cdots \ga _{n-1}$
%are almost angles. So, it would be natural to suppose that
%the functions 
%$F(t_1, \cdots t_n,
%b_1,\cdots ,b_{n-1})$ give complete set of functions of the
%phase space and there is no need in considering additional
%functions obtained by applying $\partial _j$.
%This argument would be true if we condidered more general functions
%of $\ga _j$ (say, $ C^{\infty}$), but we restrict ourselves with polynomials.
%One can easily find an example of observable
%of the phase space which cannot be written as a polynomial of
%$t _i$ and $\ga _j$.
%Still it is true that there are much more functions of the form (\ref{fun})
%than is needed.

Assuming that the conjecture is true one realizes that the
way of presenting an element of $\CA _0$ as
a linear combination of the expressions (\ref{fun}) 
may be not unique. Indeed, let us calculate the
character of the the space span by (\ref{fun}). We prescribe the
degree $i$ to $\partial _i$ which is consistent with the Poisson
brackets (\ref{poisson}). Obviously, the character is
\debut
{1\over [n-1]!\ [n]!\ [n-1]!}>\chi (q)
\non
\fin
where $\chi (q)$ is the character (\ref{char}). So, there must be
linear dependence between the functions (\ref{fun}) which 
is responsible for differential equations on the generalized
Weierstrass functions. Moreover, there is a criterium which allows to
judge whether the set of equations is complete. Indeed, to show the
completeness of the equations one has, obviously, to prove that
taking them into account leads to correct character (\ref{char}).

Let us find the equations in question.
To this end we shall use the
Fourier transform. Consider a function $F(t_1, \cdots t_n,
b_1,\cdots ,b_{n-1})$. The variables $t_j$ are the integrals of motion
(and the moduli of the Riemann surface)
and the variables $b_j$ are the functions on the Jacobi variety due
to the equations of motion. Hence
\debut 
&&F(t_1, \cdots t_n,b_1(\tau ),\cdots ,b_{n-1}(\tau ))
=\non\\&&\hskip 1cm =
\sum\limits _{k_1,\cdots ,k_{n-1}}
e^{-i\Sigma k _j\theta _j}
\int\limits _0^{2\pi}d\theta '_{1}\cdots
\int\limits _0^{2\pi}d\theta '_{n-1}
F(t_1, \cdots t_n,
b_1(\theta'),\cdots ,b_{n-1}(\theta'))e^{i\Sigma k _j\theta '_j}
\non
\fin
where $\theta _j =\sum _l A_{jl}\tau _{n-l}$.
Let us undo the Abel transformation inside the integrals:
\debut &&F(t_1, \cdots t_n,b_1(\tau ),\cdots ,b_{n-1}(\tau ))
=\non\\&&\hskip 1cm =
{1\over \det (A)}\sum\limits _{k_1,\cdots ,k_{n-1}}
e^{-i\Sigma k _j\theta _j}
\int\limits _{a_1}{d\ga _1\over \sqrt{P(\ga _1)}}\cdots
\int\limits _{a_{n-1}}{d\ga _1\over \sqrt{P(\ga _{n-1})}}
\ \prod\limits _{i<j} (\ga _i-\ga _j)\non\\&&\hskip 1cm \times
\tilde{F}(t_1, \cdots t_n,
\ga _1,\cdots ,\ga _{n-1})
\prod\limits _{j}e^{i\Phi _k(\ga _j)}
%\exp \(i\sum _j \Phi _{k}(\ga _j)\)
\label{integr}
\fin
where $\tilde{F}(t,\ga)=F(t, b(\ga))$ (recall that $b _j$
are elementary symmetric functions of $\ga$'s),
for any \newline
$k=\{k_1,\cdots ,k_{n-1}\}$ we define
$$\Phi _{k}(\ga)=\int\limits ^{\ga}k_j\omega _j$$
Deriving (\ref{integr}) we needed to take into account the
Jacobian of the Abel transformation. Later we shall relate the
integrals in (\ref{integr}) to the quasi-classical limit of
the matrix elements in quantum Toda chain.
The equations of motion correspond to vanishing
of all the integrals in (\ref{integr}). Let us explain the 
possible reasons
for these integrals to vanish.

Consider first the term in (\ref{integr}) with $k=0$ which is nothing
but the average of $F$ over the Jacobi variety (the trajectory):
\debut
\langle\  F\ \rangle ={1\over \det (A)}
\int\limits _{a_1}{d\ga _1\over \sqrt{P(\ga _1)}}\cdots
\int\limits _{a_{n-1}}{d\ga _1\over \sqrt{P(\ga _{n-1})}}
\ \prod\limits _{i<j} (\ga _i-\ga _j)\ \tilde{F}(t,\ga)
\label{aver}
\fin
There are two reasons for this integral to vanish.
First one is obvious, it is due to  existance of exact forms.
With arbitrary polynomial $L(\ga)$ associate the polynomial
$$D_t(L)(\ga)=P(\ga ){dL(\ga )\over d\ga}+{1\over 2}{dP(\ga )
\over d\ga}L(\ga )$$
We marked explicitly the dependence on the moduli (integrals
of motion) $t=\{t_1,\cdots,t_n\}$ which enters through $P(\ga )$.

There is an obvious
\newline
{\bf Proposition 1.}
{\it The following integral vanishes}
\debut
\int\limits _{c} {1\over\sqrt {P(\ga)}}D_t(L)(\ga)=0
\label{exfo1}
\fin
{\it for any  polynomial $L$ and any closed cycle $c$.}

Hence
the integal (\ref{aver}) vanishes if
\debut
\tilde{F}(t,\ga)=\sum\limits _{i=1}^{n-1}
{1\over\prod\limits _{j\ne i}(\ga _i -\ga _j)}
\ D_t(L)(\ga _i)\ G (\ga _1, \cdots \widehat{\ga _i}\cdots ,
\ga _{n-1})
\non
\fin
for any polynomial $L$ and any symmetric polynomial of $n-2$
variables $G$ (both of them can be also polynomials
of parameters $t$).
This property means in particular that by adding exact forms
one can reduce the degree of the polynomial $\tilde{F}$ in every $\ga_j$
up to $n$. 

The second reason for the integral (\ref{aver}) to vanish is
due to Riemann bilinear identity. Consider the anti-symmetric
polynomial of two variables
\debut
C_t(\ga _1,\ga _2)=R_t(\ga _1,\ga _2)-R_t(\ga _2,\ga _1)
\label{riemann}
\fin
where
$$R_t(\ga _1,\ga _2)=\sqrt{P(\ga _1)}{d\over d\ga _1}\({1\over \ga _1-\ga _2}
\sqrt{P(\ga _1)}\)$$
For any two cycles on the Riemann surface
one has 
$$\int\limits _{c_1}\int\limits _{c_2}{1\over\sqrt{P(\ga _1)}}
{1\over\sqrt{P(\ga _1)}}C_t(\ga _1,\ga _2)=c_1 \circ c_2 $$
where $\circ$ means the intersection number. Since the
cycles $a _j$ do not intersect one has the following\newline
{\bf Proposition 2.} {\it For any two a-cycles $a_j$ and $a_k$
the following integral vanishes:}
$$\int\limits _{a_j}\int\limits _{a _k}{1\over\sqrt{P(\ga _1)}}
{1\over\sqrt{P(\ga _1)}}\ C_t(\ga _1,\ga _2)=0 $$

Hence  the integral
(\ref{aver}) vanishes if 
\debut
\tilde{F}(t,\ga)=\sum\limits _{i<j}{1\over (\ga _i -\ga _j)
\prod\limits _{l\ne i,j}(\ga _i -\ga _l)(\ga _j -\ga _l)}
\ C_t(\ga _i,\ga _j)
\ G (\ga _1, \cdots \widehat{\ga _i}\cdots \widehat{\ga _j}
\cdots ,\ga _{n-1})
\non\fin

Let us consider now the case of arbitrary $k=\{k_1,\cdots ,k_{n-1}\}$.
Introduce the polynomials $S_{t,k}$:
$$i\sum\limits _{j=i}^{n-1}k_j\ \omega _j(\ga) =
{S_{t,k}(\ga)\over\sqrt{P(\ga)}}d\ga $$
Integrating by parts one gets the following three simple propositions
\newline
{\bf Proposition 1'.}{\it For any polynomial $L$ define the polynomial}
$$D_{t,k}(L)(\ga )=D_t(L)(\ga )-S_{t,k}(\ga)\int\limits _{0}^{\ga}
L(\ga ')S_{t,k}(\ga ')d\ga '$$
{\it then the following integral vanishes for any a-cycle:}
\debut
\int\limits _{a _j}{d\ga\over\sqrt{P(\ga)}}e^{i\Phi _k(\ga)}\ D_{t,k}(L)(\ga)=0
\non
\fin
{\bf Proposition 2'.}{\it Define the polynomial}
\debut
C_{t,k}(\ga _1,\ga _2)=
C_t(\ga _1,\ga _2)-
S _k(\ga _1)\int\limits _{0}^{\ga _1}
{S_{t,k}(\ga ) -S_{t,k} (\ga _2)\over \ga -\ga _2}d\ga+
S _k(\ga _2)\int\limits _{0}^{\ga _2}
{S_{t,k}(\ga ) -S_{t,k} (\ga _1)\over \ga -\ga _1}\d\ga 
\non
\fin
{\it then for any two a-cycles $a_j$ 
and $a _k$ the following integral vanishes:}
\debut
&&\int\limits _{a _j}{d\ga _1\over\sqrt{P(\ga _1)}}
\int\limits _{a _k}{d\ga _1\over\sqrt{P(\ga _2)}}e^{i\Phi _k(\ga_2)}
e^{i\Phi _k(\ga_1)}\ C_{t,k}(\ga _1,\ga _2)=0
\non
\fin   
There is one more identity which is trivial in the case $k=0$:
\newline
{\bf Proposition 3'.}{\it For any a-cycle the following integral vanishes:}
\debut
\int\limits _{a _j}{d\ga\over\sqrt{P(\ga)}}e^{i\Phi _k(\ga_2)}
\ S_{t,k}(\ga)=0
\non
\fin

From these three Propositions one finds the following
partial differential equations on the symmetric functions
of $\ga $:
\newline
1. For any polynomial one variable $L$ and any symmetric polynomial
of $n-2$ variables $G$ the equation holds:
\debut
&&\sum\limits _{i=1}^{n-1}
{1\over\prod\limits _{j\ne i}(\ga _i -\ga _j)}
D_t(L)(\ga _i)\ G (\ga _1, \cdots \widehat{\ga _i}\cdots ,
\ga _{n-1})-\non\\
&&-\sum\limits _ {l,m=1}^{n-1}\partial _l\partial _m\[
\sum\limits _{i=1}^{n-1}
{1\over\prod\limits _{j\ne i}(\ga _i -\ga _j)}
\ga _i ^{n-1-l}\int\limits _{0}^{\ga _i}L(\ga)\ga  ^{n-1-m}d\ga
\ G (\ga _1, \cdots \widehat{\ga _i}\cdots ,
\ga _{n-1})\]=0
\label{exfo}
\fin
2. For any symmetric polynomial of $n-3$ variables $G$ the equation
holds
\debut
&&\CC (G)\equiv
\sum\limits _{i<j}{1\over (\ga _i -\ga _j)
\prod\limits _{l\ne i,j}(\ga _i -\ga _l)
(\ga _j -\ga _l)}
\ C_t(\ga _i,\ga _j)
\ G (\ga _1, \cdots \widehat{\ga _i}\cdots \widehat{\ga _j}\cdots ,
\ga _{n-1})-\non\\&&\hskip 0.5cm -
\sum\limits _ {l,m=1}^{n-1}\partial _l\partial _m\[
\sum\limits _{i<j}{1\over (\ga _i -\ga _j)
\prod\limits _{l\ne i,j}(\ga _i -\ga _l)(\ga _j -\ga _l)}
\ G (\ga _1, \cdots \widehat{\ga _i}\cdots \widehat{\ga _j}\cdots ,
\ga _{n-1})
\right.\non\\ 
&&\hskip 0.5cm \times\(\left.
\ga _i^{n-l-1}\int\limits _{0}^{\ga _i}
{\ga ^{n-1-m}-\ga _j ^{n-1-m}\over \ga -\ga _j}d\ga -
\ga _j^{n-l-1}\int\limits _{0}^{\ga _j}
{\ga ^{n-1-m}-\ga _i ^{n-1-m}\over \ga -\ga _i}d\ga \)
\]=0
\label{C}
\fin
3. For any symmetric polynomial of $n-2$ variables $G$ the
equation holds:
\debut
\CQ (G)\equiv \sum\limits _{l=1}^{n-1}\partial _l\[
\sum\limits _{i=1}^{n-1}
{1\over\prod\limits _{j\ne i}(\ga _i -\ga _j)}
\ \ga _i ^{n-l-1}
G (\ga _1, \cdots \widehat{\ga _i}\cdots ,
\ga _{n-1})\]=0
\label{Q}
\fin
One must add to the equations (\ref{exfo}), (\ref{C}), (\ref{Q}) their trivial
consequences i.e. the equations obtained from them by applying
arbitrary number of $\partial _i$. We claim that in this way all
the equations of motion can be described. 

Let us illustrate this point 
considering the simple example $n=2$. In that case genus is
equal to $1$ and we have only one variable $\ga$. The equations
(\ref{C}),(\ref{Q}) are trivial, so, we are left with (\ref{exfo}).
Let us take $L(\ga)=\ga ^p$, $p=0,1,\cdots$. Then (\ref{exfo})
turns into
\debut
p\ga ^{p-1}\ P(\ga)+{1\over 2}\ga ^p\ {d\over d\ga}P(\ga)=
\partial _1^2 \({1\over p+1}\ga ^p\),
\label{wei}
\fin
we have only one time $\tau _1$ in that case.
As it has been said earlier the function $\ga$ is the
Weierstras $\CP$-function.
The equation (\ref{wei}) coincides with the usual equation (\ref{weier})
when $p=0$. For other $p$ we get the equations on
degrees of $\ga $ which can be verified for the Weierstrass function.
Recall that we were considering the space of functions of the
kind
\debut 
\partial _1^k F(t_1,t_2, \ga)
\label{xx}
\fin
Let us calculate the character of this space modulo the equations
(\ref{wei}) and their trivial consequences (those obtained
by application of $\partial _1$).
Obviously, using (\ref{wei}) we can express any function  
of the kind (\ref{xx}) as linear combination of 
$$F_0(t_1,t_2),\ F_1(t_1,t_2)\partial _1^m \ga ,
\ F_2(t_1,t_2)\partial _1^m \ga ^2 $$
So, the character is
$${1\over [2]!}\(1+{q\over [1]}+{q ^2\over [1]}\)={1\over [2]!}(1+q^2)$$
which coincides with (\ref{char}).

Let us consider the general case. We have the space of symmetric
functins of $\ga _1,\cdots ,\ga _{n-1}$ with coefficients
in $t_1,\cdots ,t_n$ on which the derivatives $\partial _1,\cdots ,
\partial _{n-1}$ act. We can consider the derivatives
also as coefficients identifying this space with $H _{n-1}$
which is the space of symmetric polynomials of
$\ga _1,\cdots ,\ga _{n-1}$ with coefficients in 
$t_1,\cdots ,t_n$ and $\partial _1,\cdots ,
\partial _{n-1}$ (recall that $\partial _i $ and $t_j$ commute).
Substracting the exact forms (\ref{exfo}) we finish with the
space $\widehat{H} _{n-1}$ which is the subspace of $H _{n-1}$
defined by the condition that the degree of the polynomials in
every $\ga _j$ does not exceed $n$. We define the spaces
$\widehat{H}_j$ ($j\le n-1$) 
of symmetric polynomials of $j$ variables $\ga _i$
whose degree in every variable does not exceed $2n-j-1$ with
coefficients in $t_1,\cdots ,t_n$ and $\partial _1,\cdots ,
\partial _{n-1}$. The action of the operators $\CC$ and $\CQ$
defined by (\ref{C}),(\ref{Q})
can be obviously extended to the action from $H_{n-3}$ to $H_{n-1}$ and 
from $H_{n-2}$ to $H_{n-1}$ respectively. It is also clear that the
images of respectively $\widehat{H}_{n-3}$ and $\widehat{H}_{n-2}$ belong to
$\widehat{H}_{n-1}$. One can easily generalize the definitions
of $\CC$ and $\CQ$ allowing them to act respectively
from $\widehat{H}_{j-2}$ to $\widehat{H}_j$ and from
$\widehat{H}_{j-1}$ to $\widehat{H}_j$. For these operators one has
\debut
[\CC ,\CQ]=0,\qquad \CQ ^2=0, \qquad {\rm Ker}|_{
\widehat{H}_{j-2}\to \widehat{H}_j}(\CC)=0
\fin
Now noticing that ${\rm deg}(\CC )=2 $ and ${\rm deg}(\CQ )=1 $
one evaluates the character:
\debut
&&{1\over [n-1]!\ [n]!}
\left\{
\(
\[{2n-1\atop n-1}\]-q\[{2n-1\atop n-2}\]
+q^2 \[{2n-1\atop n-3}\]-\cdots +(-q)^{n-1}\[{2n-1\atop 0}\]
\)-
\right.
\non\\&&\hskip 2cm
\left.
-q^2
\(
\[{2n-1\atop n-3}\]-q\[{2n-1\atop n-4}\]
+q^2 \[{2n-1\atop n-5}\]-\cdots +(-q)^{n-3}\[{2n-1\atop 0}\]
\)
\right\}=
\non\\&&\hskip 2cm ={1\over [n]!\ [n-1]!}
\(\[ 2n-1\atop n-1\]-q\[ 2n-1\atop n-2\]\)\non\fin
which coincides with the character (\ref{char}).
There is obvious similarity of what we have done with
the paper \cite{bbs1}.

\def\h{\hbar}

\section{Quantum Toda chain }

Following Sklyanin \cite{skl} we shall use the same notations
for the quantum analogues of classical objects that have been used 
in the classical case. The Hamiltonian of the quantum
periodical Toda chain is given by (\ref{ham}) with $p_j, q_j$ being
the canonical operators
$$[p_i ,q_j]=i\hbar \delta _{i,j}$$

Consider the same definition of L-operator and monodromy matrix
as in classics. The monodromy matrix satisfies the relations
\debut
R(\la -\mu)(M(\la )\otimes M(\mu ))=
(M(\mu )\otimes I)(M(\la )\otimes I)R(\la -\mu)
\label{RTT}
\fin
with $R(\la)$ being the quantum R-matrix:
$$R(\la)=\la -i\h P$$
The trace of the monodromy matrix provides $n$ commutative
integrals of motion.The center 
of the algebra is created by the quantum determinant
\debut 
A(\la )D(\la +i\h )-B(\la )C(\la +i\h )=1
\label{q-det}
\fin

The idea of using the separated variables in quantum case
goes back to \cite{gutz}. It was developed as universal method by
Sklyanin.
Let us briefly review the method of separation of variables
following \cite{skl}. From the relations
(\ref{RTT}) one finds, in particular, that
$$[B(\la),B(\mu)]=0$$
So, presenting the operator $B(\la )$ in the form
$$B(\la )=b\prod\limits _{j=1}^{n-1}(\la -\ga _j)$$
one gets a family of commuting operators:
$$[\ b,\ga _j]=0,\qquad [\ga _i,\ga _j]=0$$
As it was in the classical case the operators $b$ and $a_1 =P$ commute
with everything except between themselves:
$$[a _1, b]=i\h b$$

The idea of the method of separation of variables is in considering the
model in $b, \ga $-representation. The canonically conjugated operator to $b$
exists already: this is $a_1$. The canonically conjugated operators
for $\ga _j$ are constructed as follows.
Consider the operators
\debut
&&A(\la )=\la ^n +\la ^{n-1}a_1 +\cdots +a_n\non\\
&&D(\la )=\la ^{n-2}d _2 +\cdots +d_n
\non
\fin
then it can be shown that the operators
\debut
&&\tilde{\La} _j =\ga _j ^n +\ga _j ^{n-1}a_1 +\cdots +a_n\non\\
&&\La _j =\ga _j ^{n-2}d _2 +\cdots +d_n
\label{Lambda}
\fin 
satisfy
\debut
\tilde{\La} _j \La _j=1,\qquad [\La _j,\ga _g]=i\h\delta _{j,k}\La _j
\label{commu}
\fin
The order of multipliers in (\ref{Lambda}) is important.

It is possible to reconstruct the operators $A(\la ),B(\la ),C(\la ),D(\la )$
using $a_1, b, \ga _j, \La _j$. In particular,
\debut
T(\la)=\sum\limits _{k=1}^{n-1}\prod\limits _{j\ne k}
\({\la -\ga _j\over \ga _k -\ga _j}\)\ \(\La _k +\La _k ^{-1}\)+
\prod\limits _{j\ne 1}^{n-1}(\la -\ga _j)\(\la +a_1+\sum \ga _j\)
\label{trace}
\fin
The Hilbert space splits into a direct sum of orthogonal
subspaces $H _p$ corresponding to different
eigenvalues $p$ of the zero-mode $a _1$. 
Let us consider the space $H _0$, the solution in other 
subspaces $H_p$
are obtained from the one for $H_0$ by simple transformation.
In $H_0$ the eigenfunctions of $T(\la)$ 
with the eigenvalue $t(\la )$ 
in $\ga$-representation
can be looked for in the form
\debut
\langle\  t\ |\ \ga \ \rangle =\prod\limits _{j=1}^{n-1}Q(\ga _j)
\non
\fin
Applying (\ref{trace}) one finds with the following equation
for $Q(\ga )$:
\debut
t(\ga)Q(\ga )=Q(\ga +i\h )+Q(\ga -i\h )
\label{tq=q+q}
\fin
where $t(\la)$ is the eigenvalue of $T(\la)$ on $\ |\ t \ \rangle$.
In the subspace $H_0$ this eigenvalue must be a polynomial of the
kind
$$t(\la )=\la ^n +O(\la ^{n-2}) $$
The equation (\ref{tq=q+q}) is one equation for two unknowns: $t$ and $Q$,
so, at the first glance it is rather useless. However, assuming
certain analytical properties of $Q(\ga )$ this single
equation actually defines the spectrum. Namely, require that 
the function $Q(\ga )$ is an entire function of $\ga$ with
infinitely many real zeros. Moreover, impose the following asymptotic:
\debut
&&Q(\la)\sim \cos \( {\la n \over\h}\log \({\la\over e}\)+{\pi\over 4}\),
\qquad \la\sim\infty,
\non\\
&&Q(\la)\sim e^{{\pi\over\h }\la n}
\cos \( {\la n \over\h}\log \(-{\la\over e}\)+{\pi\over 4}
\),\qquad \la\sim -\infty
\label{asq}
\fin
The normalization of $Q$ is not the same as in \cite{skl}.
For $n=4k$ the function $Q$ differs from the function
$\varphi $ from \cite{skl} by the multiplier $\exp(\pi\la n/2\h)$
which is in this case a quasi-constant: does not disturb the equations
(\ref{tq=q+q}). If $n\ne 4k $ there is essential discrepancy with 
\cite{skl}, however, we claim that these are (\ref{asq}) which
describe correct asymptotical behaviour. 
Our normalization basically coincides with the one accepted in
\cite{gp}.
According to \cite{gp} it provides the only way to
have correct quasi-classical limit. We are going to discuss this limit in the
next section.

The main conjecture of the paper \cite{skl} is that the spectrum of
the model is defined by all the solutions to the equations (\ref{tq=q+q})
with the analytical properties of $t$ and $Q$ described above and
the asymptotic behaviour of $Q$ given by the formulae (\ref{asq}).
This conjecture was proven by Gaudin and Pasquier \cite{gp}.

Now we want to discuss the
properties of the matrix elements of the operators.
To consider the matrix elements we need to know the
scalar product in the space of functions of $\ga _j$. This scalar
product
was found by Sklyanin \cite{skl}, let us repeat the essential
steps because again there will be a minor difference
if $n\ne 4k$. Consider an operator $\CO$ which is given
by a symmetric function $F(\ga _1,\cdots ,\ga _{n-1})$. The wave-functions
are real, so, the matrix element is given by
\debut
\langle\  t \ |\ \CO\ |\  t '\ \rangle=
\int\limits _{-\infty}^{\infty}d\ga _{1}\cdots
\int\limits _{-\infty}^{\infty}d\ga _{n-1}\prod\limits _{j=1}^{n-1}
Q(\ga _j)Q'(\ga _j)F(\ga _1,\cdots ,\ga _{n-1})w(\ga _1,\cdots \ga _{n-1})
\non
\fin
where $w$ is certain weight.
Requiring that the operator $T(\la)$ for real $\la$ is self-adjoint
and using the formula (\ref{trace}) one concludes that
\debut
w(\ga _1,\cdots \ga _{n-1})=\prod\limits _{i<j}(\ga _i-\ga _{j})
\tilde{w}(\ga _1,\cdots \ga _{n-1})
\non
\fin
where $\tilde{w}(\ga _1,\cdots \ga _{n-1})$ is $i\h$-periodic entire
function of its arguments. There are two formal reasons for choosing
particular $\tilde{w}$. First, everything under
the integral is symmetric with respect to $\ga _1,\cdots \ga _{n-1}$
except for the multiplier $\prod _{i<j}(\ga _i-\ga _{j})$ in $w$, so
it does not make sense to put in $\tilde{w}$ anything that can be 
killed by anti-symmetrization. Second, requiring convergence of the
integrals one finds from the asymptotic (\ref{asq}) that $\tilde{w}$
can contain $\exp(-2\pi\ga _jm/\h)$ with $1\le m\le n-1$ . These
two
requirements fix $\tilde{w}$ up to anti-symmetrization:
\debut
\tilde{w}(\ga _1,\cdots \ga _{n-1})=\prod\limits _{j=1}^{n-1}
e^{{2\pi\over\h}\ga _j (j-n)}
\label{ww}
\fin
Again, if $n=4k$ this essentially coincides with \cite{skl},
otherwise there is a minor discrepancy.

Informal reason for taking $\tilde{w}$ in the form (\ref{ww})
refers to the quasi-classics which will be discussed in the
next section.

Similarly to the classical case the algebra of observables $\CA _0$
is defined as the algebra generated by all the
coefficients of $A(\la )$, 
$\tilde{B}(\la )=b ^{-1}B(\la )$, $\tilde{C}(\la )=b\ C(\la )$, 
$D(\la )$. The commutation relations
(\ref{RTT}) and the quantum determinant (\ref{q-det}) provide
sufficiently many relations to show that the quantum algebra of 
observables has the same size as the classical one.
To make this statement mathematically
rigorous one says that their characters coincide.

Now we formulate an analogue of the Conjecture 1 of the previous section.
\newline
{\bf Conjecture 2.}{\it Every quantum observable $\CO$ can be presented in the
form}
\debut
\CO =G_L(t_1,\cdots ,t_n)F(b _1,\cdots ,b_{n-1})G_R(t_1,\cdots ,t_n)
\non
\fin
{\it where $G_L$, $F$, $G_R$ are polynomials.}
\newline
This conjecture looks more natural than its classical counterpart and
explains the mystery of the latter. The point is that there is no
closed formula for the commutation relations of $T(\la)$ and $B(\mu)$
which would allow ordering of the operators $t_i$ and $b_j$.

Taking the matrix element between two eigenstates of the Hamiltonians
one can essentially neglect the polynomials $G_L$ and $G_R$ because
acting on the eigenstates they produce the eigenvalues. Hence we are 
mostly interested in the matrix elements of the operators $\CO _0=
F(b _1,\cdots ,b_{n-1})$ which are given by 
the integrals of the king
\debut
\langle\  t \ |\ \CO _0\ |\  t'\ \rangle=
\int\limits _{-\infty}^{\infty}d\ga _{1}\cdots
\int\limits _{-\infty}^{\infty}d\ga _{n-1}\prod\limits _{j=1}^{n-1}
Q(\ga _j)Q'(\ga _j)
\prod\limits _{i<j}(\ga _i-\ga _{j})\prod\limits _{j=1}^{n-1}
\tilde{F}(\ga _1,\cdots ,\ga _{n-1})
\prod\limits _{j=1}^{n-1}e^{{2\pi\over\h} \ga _j (j-n)}
\label{me}
\fin
where
\debut
\tilde{F}(\ga _1,\cdots ,\ga _{n-1})=F(b _1(\ga ),\cdots ,b_{n-1}(\ga ))
\label{fff}
\fin
The function
$$\prod\limits _{i<j}(\ga _i-\ga _{j})\prod\limits _{j=1}^{n-1}
\tilde{f}(\ga _1,\cdots ,\ga _{n-1})$$
is an anti-symmetric polynomial of $\ga _1, \cdots ,\ga _{n-1}$. Every anti-symmetric
polynomial can be presented as a linear combination of Schur functions, so, 
without loss of generality we can replace (\ref{fff}) by
$$\det \(F_i(\ga _j)\)$$
for some polynomials of one variable $F_1,\cdots ,F_{n-1}$.
Then the matrix elememt can be presented as a determinant of 
$(n-1)\times (n-1)$-
matrix composed of
one-fold integrals:
\debut
\det \(\int\limits _{-\infty}^{\infty} Q(\ga )Q'(\ga )F _i(\ga)
e^{{2\pi\over\h} \ga  (j-n)}\)d\ga 
\label{det}
\fin
We consider the integral
\debut
\int\limits _{-\infty}^{\infty} Q(\ga )Q'(\ga )F (\ga)
e^{-{2\pi\over\h} \ga  k}d\ga
\label{qab}
\fin
as a deformation of hyper-elliptic integral. The study of the properties of these
integrals is very important for understanding of matrix elements
of operators in the model.

Since the quantum algebra of observables has the same character as the 
classical one the equations (\ref{exfo}), (\ref{C}), (\ref{Q}) must have their
quantum counterparts. Hence there must be identities for the integrals
(\ref{qab}) from which these quantum counterparts follow.
These identities can indeed be found, we are going to describe them in the same
order as it has been done in classical case.

Let us introduce the operation $\Delta$ which attaches to every
function $F(\ga )$ the function
$$\Delta (F)(\ga )=F(\ga +i\h )-F(\ga -i\h )$$
The operation $\Delta ^{-1}$ is not always defined, but on
polynomials it is. For any polynomial $L$ one can define a
polynomial $F$ such that $F(\ga )=\Delta ^{-1}(L)(\ga )$. For uniqueness we
require also that $\Delta ^{-1}(L)(0)=0$.
Consider the integral (\ref{qab}) with $Q(\ga)$, $Q'(\ga)$
satisfying the equations (\ref{tq=q+q}) with the eigenvalues
$t(\ga)$ and $t'(\ga)$.
For any given polynomial $L(\la)$ construct the polynomial
\debut
&&D(L)_{t,t'}(\ga)=
t(\ga )\Delta ^{-1}(Lt)(\ga )+t'(\ga )\Delta ^{-1}(Lt')(\ga )-\non\\
&&\hskip 2cm-
t(\ga )\Delta ^{-1}(Lt')(\ga -i\h  )-
t'(\ga )\Delta ^{-1}(Lt)(\ga -i\h  )-\non\\&&\hskip 2cm-
L(\ga)t(\ga)t'(\ga) +L(\ga +i\h )-L(\ga -i\h )
\label{qexfo}
\fin
Then we have the following analogue of the Propositions 1 and 1':
\newline
{\bf Proposition 1''.}{\it For any $1\le k\le n-1$ the following
integral vanishes:}
\debut
\int\limits _{-\infty}^{\infty} Q(\ga )Q'(\ga )D(L)_{t,t'}(\ga)
e^{-{2\pi\over\h} \ga  k}d\ga=0
\label{qclosed}
\fin

Using these q-exact forms (\ref{qexfo}) we can always reduce the degree of
the polynomial $F(\ga )$ under the integral (\ref{qab}) in order that
it does not exceed $2n-2$. So, we are left with finite number
of different integrals (\ref{qab}) with $F(\ga )=\ga ^j$,
$j=0,1,\cdots ,2n-2 $. These integrals are subject to further
relations.

In order to define 
a quantum analogue of the polynomial $C_t(\ga _1, \ga _2)$ (\ref{riemann}).
we shall need some preparations.
Consider the function 
$$U(\ga ,\delta)={t(\ga )-t(\delta)\over\ga -\delta} $$
The notation $\Delta ^{-1}(U(\cdot ,\delta ))(\ga)$ means
that $\Delta ^{-1}$ is applied to the first argument, i.e.
$$\Delta(\Delta ^{-1}(U(\cdot ,\delta )))(\ga)=U(\ga ,\delta)$$
The function $U'$ is defined in the same way replacing $t$ by $t'$.

Consider the anti-symmetric polynomial of two variables
$$C_{t,t'}(\ga _1 ,\ga _2)=R_{t,t'}(\ga _1 ,\ga _2)-R_{t,t'}(\ga _2 ,\ga _1)$$
where $R_{t,t'}(\ga _1 ,\ga _2)$ is defined as follows:
\debut
&&R_{t,t'}(\ga _1 ,\ga _2)=t(\ga _1)\Delta ^{-1}(U(\cdot ,\ga _2 ))(\ga _1)+
t'(\ga _1)\Delta ^{-1}(U'(\cdot ,\ga _2 ))(\ga _1)-\non\\&&\hskip 2cm-
t(\ga _1)\Delta ^{-1}(U'(\cdot ,\ga _2 ))(\ga _1-i\h )-
t'(\ga _1)\Delta ^{-1}(U(\cdot ,\ga _2 ))(\ga _1-i\h )-\non\\&&
\hskip 2cm -{1\over 2}
{1\over \ga _1 -\ga _2}(t(\ga _1)-t(\ga _2))(t'(\ga _1)-t'(\ga _2))
\non
\fin
We have the \newline
{\bf Proposition 2''.}
{\it For any $1<k,l<n-1$ the following integral vanishes:}
\debut
\int\limits _{-\infty}^{\infty}\int\limits _{-\infty}^{\infty}
 Q(\ga _1 )Q'(\ga_1 ) Q(\ga _2 )Q'(\ga _2)C_{t,t'}(\ga _1,\ga _2)
e^{-{2\pi\over\h}\ga _1 k}e^{-{2\pi\over\h} \ga _2 l}d\ga _1d\ga_2 =0
\label{qC}
\fin
 
Finally, let us define
\debut
S _{t,t'}(\ga )=t(\ga )-t'(\ga )
\non
\fin
then we have\newline
{\bf Proposition 3''.}
{\it For any $1<k<n-1$ the following integral vanishes:}
\debut
\int\limits _{-\infty}^{\infty} Q(\ga )Q'(\ga ) S_{t,t'}(\ga)
e^{-{2pi\over\h} \ga  k}d\ga=0
\label{qQ}
\fin

This is in fact the simplest relation, we consider it  last  for
historical reasons. One obvious consequence of this relation is
orthogonality of the wave-function. 

Let us say a few words about proof of all these relations.
Consider the simplest one (\ref{qQ}) we have
\debut
&&Q(\ga )Q'(\ga ) S_{t,t'}(\ga)=Q(\ga )Q'(\ga )(t(\ga )-t'(\ga ))=\non\\&&
\hskip 1.2cm=
Q(\ga +i\h)Q'(\ga )+Q(\ga -i\h )Q'(\ga )
-Q(\ga )Q'(\ga +i\h)-Q(\ga )Q'(\ga-i\h )
\non
\fin
So, 
\debut
&& \int\limits _{-\infty}^{\infty} Q(\ga )Q'(\ga ) S_{t,t'}(\ga)
e^{-{2pi\over\h} \ga  k}d\ga=\non\\&&\hskip 1cm =
\int\limits _{-\infty}^{\infty}Q(\ga +i\h)Q'(\ga )
e^{-{2pi\over\h} \ga  k}d\ga
+\int\limits _{-\infty}^{\infty}Q(\ga -i\h )Q'(\ga )
e^{-{2pi\over\h} \ga  k}d\ga -\non\\&&\hskip 1cm
-\int\limits _{-\infty}^{\infty}Q(\ga )Q'(\ga +i\h )
e^{-{2pi\over\h} \ga  k}d\ga
-\int\limits _{-\infty}^{\infty}Q(\ga )Q'(\ga-i\h )
e^{-{2pi\over\h} \ga  k}d\ga
\non
\fin
by shift of contour the first integral cancels the forth one
and the second cancels the third one. The shift of contour is
possible because the function $Q$ behaves asymptotically
on the line ${\rm Im}(\ga) ={\rm const}$ essentially in the same
way as it behaves at the real axis. The relations (\ref{qclosed})
and (\ref{qC}) are proven similarly, but require more sophisticated
calculations.

Using the equations (\ref{qclosed}), (\ref{qC}), (\ref{qQ})
we can write down equations similar to (\ref{exfo}),
(\ref{C}), (\ref{Q}) which would guarantee that the number of 
operators (character of the algebra of observables)
is correct. The corresponding calculation does not differ from
the one presented earlier for the classical case.

\section{ Quasi-classical case}

We had a number of identities described by Propositions
1,1',1''; 2,2',2''; 3',3''. They look as reflection of the same
structure. The goal of this section is to explain the relation
between different levels of deformation.

Consider the quasi-classical quantization of the Toda lattice.
We had the following formulae for the symplectic form and
corresponding 1-form in the coordinates $t$ and $\ga $:
\debut 
&&\alpha =\sum\limits _{j=1}^{n-1}\log \La (\ga _j)d\ga _j\non\\
&&\omega =d\alpha =\sum\limits _{j=1}\sum\limits _{k=2}^n
{\ga _j^{n-k}\over \sqrt{ P(\ga _j)}}dt_k\wedge d\ga _j ,
\label{sympl}
\fin
it this section we shall ignore the contribution
from the center of mass variables $a_1$ and $b$ which is
easy to find if needed.
Using these formulae one immediately writes a 
quasi-classical
proposal for the
wave-function of the states with eigenvalues $t$
in $\ga$-representation:
\debut
&&\Psi _t(\ga)= \mu ^{1\over 2}\exp \({1\over i\h}
\int ^{\ga}\alpha\)=\non\\
&&\hskip 1cm =\prod\limits _{i<j}(\ga _i-\ga _j)^{1\over 2}
\prod\limits _{j=1}^{n-1} 
\({1\over P(\ga _j)}\)^{1\over 4}\exp \({1\over i\h}\int ^{\ga _j}
\log \La (\ga )d\ga \)
\label{qcwf}
\fin
where $\mu $ is defined as follows
$$\wedge ^{n-1} (\omega )=\mu \ d\ga _1\wedge\cdots\wedge d\ga _{n-1}\wedge
d t_2 \wedge\cdots\wedge dt _n$$

The relation of the formula (\ref{qcwf}) to the exact quantum formulae that
we had before is clear.\newline
1. The multiplier $\prod\limits _{i<j}(\ga _i-\ga _j)^{1\over 2}$
is a piece of the weight of integration $w$ (\ref{me}). Another one
will come from the second wave-function in the matrix elements.
The wave function without  this multiplier will be denoted by
$\tilde{\Psi }_t(\ga )$. \newline
2. The function $\tilde{\Psi }_t(\ga )$ splits into product of the
expressions
$$\({1\over P(\ga _j)}\)^{1\over 4}\exp \({1\over i\h}\int ^{\ga _j}
\log \La (\ga )d\ga \)$$
which has to be related to the quasi-classical limit of function $Q$.
One should be careful here because the function $\log \La (\ga )$ is
multi-valued, so, the branches must be defined.
Moreover, we could probably need linear combination of the
wave-functions corresponding to different branches.

Recall that 
$$\La (\ga )={1\over 2}\(T(\ga )+\sqrt {T^2(\ga )-4}\)$$
where $T(\la )$ is a polynomial of degree $n$. The polynomial
$P(\ga )=T^2(\ga )-4$ has $2n$ real simple roots $\la _1<\cdots <\la _{2n}$.
We define the function $\La (\ga )$ on the plane with the cuts along the
intervals; $I_0=(-\infty ,\la _1]$, $I _1=[\la _2,\la _3]$, ...,
$I_n =[\la _{2n}, \infty )$.
Require that $\log \La (\ga \pm i0)$ are real for $\ga >\la _{2n}$
then, obviously,
$$\log \La (\ga + i0)+\log \La (\ga - i0)=0,\qquad \ga\in I_n$$
Continuing analytically $\log \La (\ga)$ into the plane with the
cuts one finds that
$$
\bar{\log \La (\ga )}=\log \La (\bar{\ga}),\ {\rm and}\quad
\log \La (\ga + i0)+\log \La (\ga - i0)=2\pi i (n-j)
,\qquad \ga\in I_j$$

The variable $\ga _j$ belongs classically to $I_j$. So, we need the
wave function real on entire real axis and not containing the factors
$\exp (-{1\over\h}s)$ with real $s$ when all the variables are
in the classically permited places.
This requires, first,
taking different brances of $\log \La $ for different
$\ga _j$ and, second, taking a sum of two wave functions with $\ga _j +i0$
and $\ga _j-i0$ for every $\ga _j$. The result is
\debut
\tilde{\Psi }_t(\ga )=\prod\limits e^{{\pi\over \h}(j-n)\ga _j}
Q_{qc}(\ga _j)
\non
\fin
where 
\debut
Q_{qc}(\ga )=V (\ga +i0)+V(\ga -i0),
\qquad
V (\ga )=\(-{1\over P(\ga )}\)^{1\over 4}
\exp \({1\over i\h}\int ^{\ga}\log \La(\ga )d\ga\)
\non\fin
and the branch of $\log \La $ is defined above, 
$\(-{1\over P(\ga )}\)^{1\over 4}$ is real positive for $\la _{2j-1}<\ga
<\la _{2j}$.

To ensure that $Q_{qc}(\ga )$ is a single-valued function
in the plane with the cuts the Bohr-Sommerfeld quantization
condition must hold
\debut
J_j=\int\limits _{a _j}\log \La(\ga )d\ga =\pi\h (2n _j +1)
\label{bs}
\fin
the integrals $J _j$ are the classical actions. 
The cuts of $Q _{qc}$ come from condensation of zeros of
the quantum $Q$ in the quasi-classical limit.

Let us consider now the quasi-classical limit $\h\to 0$ of the
matrix elements. This limit makes sense literally if two
condition are satisfied:
\newline
1. The quantum numbers are large. In our case it means that
the zones $[\la _{2j}, \la _{2j+1}]$ do not collapse.
\newline
2. The states $\ |\  t\ \rangle $ and $\ |\  t'\ \rangle $ are close,
i.e. the eigenvalues of Hamiltonians are close:
$t _j'-t_j =O(\h )$.

Consider the matrix element (\ref{me}) for such close states.
It consists of the integrals
$$\int\limits _{-\infty}^{\infty}Q(\ga)Q'(\ga)F_i(\ga)
e^{{2\pi\over\h}(j-n)\ga}d\ga $$
From the quasi-classical estimation of $Q$, $Q'$ one concludes
that 
\debut 
&&\int\limits _{-\infty}^{\infty}Q(\ga)Q'(\ga)F_i(\ga)
e^{{2\pi\over\h}(j-n)\ga}d\ga
\ {\buildrel \h\to 0 \over\longrightarrow } \non\\
&& 4\int\limits _{I_j}{1\over\sqrt{P(\ga )}}\cos \({1\over \h}
\int ^{\ga}{\rm Re}(\log\La (\ga '+i0 ))d\ga '+{\pi \over 4}\)
\cos \({1\over \h}
\int ^{\ga}{\rm Re}(\log\La '(\ga ' +i0 ))d\ga '+{\pi \over 4}\)
F _i(\ga )d\ga 
\non
\fin
We have
\debut
&&2\cos \({1\over \h}
\int ^{\ga}{\rm Re}(\log\La (\ga  '+i0) )d\ga '+{\pi \over 4}\)
\cos \({1\over \h}
\int ^{\ga}{\rm Re}(\log\La '(\ga '+i0 ))d\ga '+{\pi \over 4}\)=\non\\&&
\hskip 3cm
=
\cos 
\({1\over \h}
\int ^{\ga}{\rm Re}(\log\La '(\ga '+i0  )-\log\La (\ga '+i0  ))d\ga '\)-
\non\\&&\hskip 3cm
-\sin 
\({1\over \h}
\int ^{\ga}{\rm Re}(\log\La '(\ga '+i0  )+\log\La (\ga '+i0  ))d\ga '\)
\non
\fin
The second term in the RHS can be thrown away because it is
rapidly oscillating in the classical limit.
The variation $\log\La (\ga )-\log\La '(\ga )$ is estimated as follows.
Consider the classical solution with $T _{cl}(\la )=t(\la )$ where 
$t(\la )$ is one of the eigenvalues. This is the place where
using the same notations for classical and quantum observables
can be misleading, so, we mark explicitly the classical
ones. The eigenvalue $t'(\la)=
T_{cl}(\la )+\delta T(\la)$. One has (in further calculations
we neglect contributions of order $o(\h )$):
$$\delta \log \La(\ga  )\equiv \log\La '(\ga )-\log\La (\ga )=
{\delta T(\ga)\over \sqrt{P(\ga )}}$$
How to find $\delta T(\ga)$? The quasi-classical states are subject 
to the Bohr-Sommerfeld quantization conditions (\ref{bs}).
The quantum numbers $n_j$ are quasi-classically large:
$n_j=O(\h ^{-1})$, but their differences for the close states
are finite $k_j\equiv n'_j-n_j=O(1)$. Hence
\debut
\delta \int\limits _{a_j}\log \La (\ga )d\ga =
\int\limits _{a_j}{\delta T(\ga)\over \sqrt{P(\ga )}}d\ga =\sum\limits
_{l=1}^{n-1}\delta t_{n-l+1}\int\limits _{a_j}
{\ga ^{l-1}\over \sqrt{P(\ga )}}d\ga=\h k_j
\fin
The matrix
$$A^{-1}_{lj}=\int\limits _{a_j}
{\ga ^{l-1}\over \sqrt{P(\ga )}}d\ga
$$
is the inverse for the matrix used in definition of the normalized
Abelian differentials $\omega _j$. Hence
$$\delta t_{n-l+1}=\h k_j A_{jl}$$
which means that
$$\delta \log \La(\ga  )d\ga =\h \sum k_j\omega _j$$
Thus the quasi-classical matrix element for the close states
is
\debut 
\langle\  t\ |\ \CO\ |\ t'\ \rangle =
\int\limits _{I_1}{d\ga _1\over \sqrt{P(\ga _1)}}\cdots
\int\limits _{I_{n-1}}{d\ga _1\over \sqrt{P(\ga _{n-1})}}
\ \prod\limits _{i<j} (\ga _i-\ga _j)
F(\ga _1,\cdots ,\ga _{n-1})
\prod\limits _{j}2\cos (\Phi _k(\ga _j))
\label{sqme}
\fin
recall the notation $\Phi _k(\ga)=\int ^{\ga }k_j\omega _j$.
This is  the same expression as for the Fourier coefficient
in (\ref{integr}). 
At the first glance
there are two disagreements: in (\ref{integr})
we integrate over $a_j$, and we have $\exp (i\Phi _k(\ga _j))$ under
the integral.
Actually, these
two disagreements compensate each other because $a_j=(I_j+i0)-(I_j-i0)$
and $\Phi _k(\ga +i0)+\Phi _k(\ga -i0)=0$ when $\ga \in I_j$.
Notice that the states $|\ t\ \rangle $ are not normalized.

Certainly it is not a wonder that we have found
the Fourier coefficient as the quasi-classical limit of the
matrix element. We have performed all the calculations in order
to have complete mathematical picture. On the other hand from
the point of view of physics one can argue as follows.

Consider the action-angle variables $J _1, \cdots , J_{n-1}$,
$\theta _1, \cdots ,\theta _j$. The Bohr-Sommerfeld
quantization in this variables does not give correct quantum
result, but still it is correct quasi-classically.
Consider the eigenstate of Hamiltonians $\ |\ t\ \rangle$ and the 
eigenstates of the angles $\ |\ \theta\ \rangle$.
Quasi-classically one has for the wave-function:
\debut
\langle\  t\ |\ \theta\ \rangle ={1\over\sqrt{\prod J_k}}\exp \({1\over i\h}
\sum
J_k\theta _k\)
\label{bswf}
\fin
Consider an operator $\CO $. This operator can be, at least quasi-classically,
ordered in such a way that $J$'s are to the left of $\theta$'s.
Then the classical shape of the corresponding observable
on the solution with given values of integral ($t$) is
$$\CO _{cl}(\theta _1,\cdots ,\theta _{n-1})=\lim _{\h\to 0}
{\langle\  t\ |\ \CO\ |\ \theta\ \rangle\over\langle\  t\ |\ \theta\ \rangle}$$
Insert the complete set of eigenstates into this formula
\debut
\CO _{cl}(\theta _1,\cdots ,\theta _{n-1})=
\lim _{\h\to 0}
{\langle\  t\ |\ \CO\ |\ \theta\ \rangle\over\langle\  t\ |\ \theta\ \rangle}=
\sum\limits _{t'}
{\langle\  t\ |\ \CO\ |\ t'\ \rangle\over
\langle\  t'\ |\ t'\ \rangle}
 {\langle\  t'\ |\ \theta\ \rangle\over\langle\  t\ |\ \theta\ \rangle}
\label{jj}
\fin
Quasi-classically only close states are important for which we
have from (\ref{bswf}):
$$
{\langle\  t'\ |\ \theta\ \rangle\over\langle\  t\ |\ \theta\ \rangle}=
e^{-i\Sigma k_j\theta _j}
$$
Now it is obvious that (\ref{jj}) gives the Fourier transformation
of $\CO _{cl}(\theta _1,\cdots ,\theta _{n-1})$. It is clear from
the equation (\ref{sqme}) that quasi-classically
$$
\langle\  t'\ |\ t'\ \rangle=\langle\  t\ |\ t\ \rangle +O(\hbar ),\qquad
\langle\  t\ |\ t\ \rangle=\det (A)$$ 
so, there is complete agreement with the formula (\ref{integr}).

Let us consider two close states introducing the notation:
$S_{t,k}(\ga )=t'(\ga )-t(\ga )$. The quantum matrix element
goes to the classical Fourier coefficient when $\h\to 0$.
If we consider the classical Fourier coefficient with
$k=0$ it describes the classical limit of the quantum
expectation value $\langle\ t \ |\CO\ |\ t \rangle $.
So, there is no wonder that we have the following sequences:
\debut
\matrix{
D_{t,t'}(L)(\ga )\ &{\buildrel \h\to 0 \over\longrightarrow }&
\ D_{t,k}(L)(\ga )
\ &{\buildrel k= 0 \over\longrightarrow }&\ D_t(L)(\ga )\non\\
C_{t,t'}(\ga _1,\ga _2 )\ &{\buildrel \h\to 0 \over\longrightarrow }&
\ C_{t,k}
(\ga _1,\ga _2 )\ &{\buildrel k= 0 \over\longrightarrow }&
\ C_t(\ga _1,\ga _2 )
\non\\
S_{t,t'}(\ga )\ &{\buildrel \h\to 0 \over\longrightarrow }&
\ S_{t,k}(\ga )
\ &{\buildrel k= 0 \over\longrightarrow }&\ 0}
\fin
So, we have two levels of deformation of hyper-elliptic differentials.
The impression is that the quantum deformation is very
natural, and that the classical mechanics appears as a strange
intermediate case.

\section{Conclusions}

The identities (\ref{qclosed}), (\ref{qC}), (\ref{qQ}) present the
main result of this paper. The show that the matrix elements
of arbitrary operator in quantum Toda chain can
be expressed with help of finitely
many integrals which possess remarkable properties.

There is a difference with what we have in Sine-Gordon
theory in infinite volume. Indeed, the matrix elements for
Toda chain are given by integrals of arbitrary deformed
differentials with respect to fixed half-basis of deformed
cycles. In Sine-Gordon case we took arbitrary half-basis.
The reason for that is the difference in the type of reality conditions.

In this connection it would be very important to consider the deformation
of Toda chain with trigonometric R-matrix
and more complicated reality conditions which is more
close to Sine-Gordon case. We hope that in this situation
there will be  a complete duality between deformed differentials
and deformed cycles.
\newline
{\bf Acknowledgments.} I am grateful to O. Babelon,
D. Bernard, L. Faddeev, E. Frenkel and N. Reshetikhin for discussions.
I would like to thank RIMS at Kyoto University where the
work was finished for hospitality.

\end{document}